# Perfect proton selectivity in ion transport through two-dimensional crystals

L. Mogg[1,2], S. Zhang[2,3], G.-P. Hao[2,4], K. Gopinadhan[2,5], D. Barry[2], B. L. Liu[6], H. M. Cheng[6], A. K. Geim[1,2], M. Lozada-Hidalgo[1,2]

[1]National Graphene Institute, The University of Manchester, Manchester M13 9PL, UK
[2]School of Physics and Astronomy, The University of Manchester, Manchester M13 9PL, UK
[3]Key Laboratory for Green Chemical Technology of Ministry of Education, School of Chemical Engineering and Technology, Tianjin University, Tianjin 300072, China
[4]State Key Laboratory of Fine Chemicals, School of Chemical Engineering, Dalian University of Technology, Dalian 116024, China
[5]Department of Physics, Indian Institute of Technology Gandhinagar, Gujarat 382355, India
[6]Shenzhen Graphene Center Tsinghua-Berkeley Shenzhen Institute, Tsinghua University 1001 Xueyuan Road, Shenzhen 518055, China

**Recent experiments have shown that defect-free monolayers of graphene and hexagonal boron nitride (hBN) are surprisingly permeable to thermal protons despite the two-dimensional (2D) materials are completely impenetrable to all gases. Even individual atomic-scale defects such as, e.g., vacancies caused notable permeation of small atoms and molecules and could be detected. However, it remains untested whether or not small ions can permeate through pristine 2D crystals. Here we show that mechanically-exfoliated graphene and hBN exhibit perfect Nernst selectivity: Only protons can permeate through these crystals, with no detectable flow of counterions. Using electrical measurements, we studied ion transport through suspended 2D membranes that had few if any atomic-scale defects as shown by gas permeation tests. The 2D crystals were then used to separate reservoirs filled with hydrochloric acid solutions. Protons accounted for all the detected current with chloride ions being blocked. These results corroborate the previous conclusion that only thermal protons are able to pierce defect-free 2D crystals. Besides the fundamental importance for understanding of the mechanism of proton transport through atomically-thin crystals, our results are also of interest for research on various separation technologies based on 2D materials.**

Proton transport through 2D crystals has recently been studied both experimentally and theoretically[1–9]. As for experiment, it was found that proton permeation through mechanically-exfoliated crystals was thermally activated with energy barriers of ≈0.8 eV for graphene and ≈0.3 eV for monolayer hBN[1]. Further measurements using deuterons, nuclei of hydrogen's isotope deuterium, showed that quantum oscillations raised the energy of incoming protons by 0.2 eV[2]. This correction yielded total barriers of ≈0.5 eV for monolayer hBN and ≈1 eV for graphene. From the theory perspective, the latter value is notably lower (by at least 30% but typically a factor of 2) than that found in density-functional calculations for graphene[3–7]. To account for the difference, a recent theory suggested that graphene can be partially hydrogenated during the measurements, which makes its lattice slightly sparser, thus more permeable to protons[8,9]. An alternative explanation put forward was to attribute the observed proton currents to atomic-scale lattice defects including vacancies[10,11]. This was argued on the basis of ion-selectivity measurements using chemical-vapor-deposited (CVD) graphene[11]. Indeed, CVD graphene is known to possess a large density of atomic-scale defects that appear during growth[12–14]. Such defects are normally absent in mechanically-exfoliated 2D crystals, which was proven most conclusively in gas leak experiments using so-called nanoballoons[15–17]. Even a single angstrom-sized vacancy per micrometer-size area could be detected in those experiments[16,17].

Whereas it is plausible that vacancies and similar defects played a dominant role in experiments using CVD graphene[10,11], extrapolation of those results to mechanically-exfoliated 2D crystals is unjustifiable. To resolve the controversy, it is crucial to carry out similar ion-selectivity measurements using mechanically-exfoliated crystals with little or no defects[1,2,15]. In this work we report such studies.

The investigated devices were fabricated using monolayer graphene and mono- and bi- layer hBN crystals that were isolated by micromechanical cleavage[18] (see Supplementary Fig. 1). The crystals were suspended over microfabricated apertures (2 μm in diameter) etched in free-standing silicon-nitride (SiN) membranes[1] (Supplementary Fig. 2). A prefabricated polymer washer with a 10 μm diameter hole was then transferred on top of the crystal so that the hole was aligned with the aperture in the SiN membrane (Supplementary Fig. 2). The assembly was baked at ~150 °C to ensure that the washer firmly clamped the 2D crystal to SiN and sealed the crystal edges in order to prevent any possible leak along the substrate. In a series of control experiments, we checked that there were no microscopic defects in our exfoliated 2D crystals by employing the approach described in Refs. [15,16] and previously also used in our experiments[1]. To this end, we made hBN and graphene membranes to cover micron-sized cavities etched in an oxidized Si wafer and tested the enclosures for possible gas leaks (see inset Fig. 1b and Supplementary section 'Leak tests using nanoballoons'). Even a single vacancy would be detectable in these measurements[16,17], but neither of the dozens of tested 2D crystals showed such leakage (Supplementary Fig. 4). In contrast, similar devices made from CVD graphene normally exhibited notable gas permeation.

The chips containing the individual 2D membranes (Supplementary Fig. 2) were then used to separate two compartments filled with hydrochloric acid (HCl) at chosen concentrations[19]. Electrical conductance through the membranes was probed using Ag/AgCl electrodes placed inside the compartments. Fig. 1a shows the current density $I$ as a function of applied voltage $V$ for representative devices made from graphene and hBN. The $I$-$V$ response was linear, which allowed us to determine the areal conductivity $\sigma = I/V$. We found monolayer hBN to be most conductive of the studied crystals, followed by bilayer hBN and monolayer graphene. For example, using 1 M HCl we found $\sigma \approx 1,000$ mS cm$^{-2}$ for monolayer hBN, ≈40 mS cm$^{-2}$ for bilayer hBN and ≈12 mS cm$^{-2}$ for monolayer graphene. The relative conductivities agree well with those found in the previous studies using Nafion (rather than HCl) as the proton-conducting medium[1]. Thicker crystals (e.g., bilayer graphene) exhibited no discernable conductance, again in agreement with the previous report[1].

Because monolayer hBN exhibited the highest conductivity, we focus our discussion below on this particular 2D material, as it allowed the most accurate ion-selectivity measurements (results for graphene are presented in Supplementary Information). Fig. 1b shows $\sigma$ found for hBN at various HCl concentrations (the same concentration was used in both compartments). For concentrations above 1 mM, $\sigma$ increased linearly with HCl concentration. At lower concentrations, the measured current was below our detection limit. The latter was determined by electrical leakage along surfaces of the liquid cell and was of the order of 1 pA as found using control devices with no holes in SiN membranes[19]. In another control experiment, we used devices with the same SiN aperture but without a 2D crystal. They exhibited conductance at least ~1000 times larger than that for the devices with graphene or hBN crystals covering the aperture (Supplementary Fig. 3). This demonstrates that the reported values of $\sigma$ were limited by the relatively low ion permeation through 2D crystals, and the series resistance due to the electrolyte itself could be neglected.

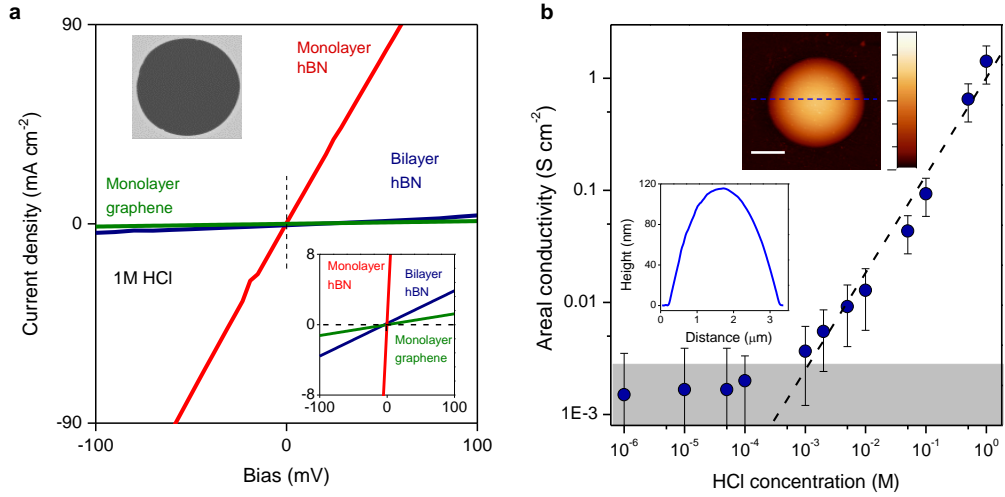

**Figure 1| Proton transport through 2D crystals studied using aqueous solutions.** (**a**) Examples of *I-V* characteristics for 1 M HCl. Bottom inset: Zoom-in. Top inset: Electron micrograph of a suspended hBN membrane (aperture diameter, 2 μm). (**b**) Concentration dependence of the areal conductivity $\sigma$ for monolayer hBN. Grey area indicates our detection limit given by parasitic leakage currents. Error bars: SD from different measurements. Dashed line: Best linear fit to the data. Top inset: atomic force microscopy (AFM) height profile of an 'inflated nanoballoon'. Here, graphene monolayer seals a micron-sized cavity containing pressurized Ar. The pressure difference across the membrane makes it to bulge up. Lateral scale bar, 1 μm; color scale, 130 nm. Bottom inset: AFM line trace taken along the blue dotted line in the top inset.

The measured conductivity could be due to either H⁺ or Cl⁻ or both ions permeating through 2D crystals. For the purpose described in the introduction, it is necessary to determine the fraction of *I* carried by each of these species. Such fractions are usually referred to as transport numbers[20] ($t_H$ and $t_{Cl}$ for protons and chloride, respectively) and, by definition, they satisfy $t_H + t_{Cl} \equiv 1$ and the inequality: $0 \leq$ both $t_H$ and $t_{Cl} \leq 1$. To find their values for our 2D membranes, we used the same setup as in the measurements discussed in Fig. 1 but with different HCl concentrations in the two compartments (inset of Fig. 2b). The concentration gradient drives both H⁺ and Cl⁻ ions towards equilibrium, from the high concentration ($C_h$) compartment to the low concentration ($C_l$) one. Therefore, the sign of the total ionic current at zero *V* indicates whether the majority carriers are protons (positive *I*) or chloride ions (negative). Fig. 2a shows typical *I-V* characteristics for monolayer hBN devices and concentration ratio $\Delta C \equiv C_h/C_l$ =10. Independently of the absolute values of HCl concentrations, the zero-*V* current was always positive proving that protons dominate ion transport through our membranes. The same behavior was found for graphene devices (Supplementary Fig. 5).

The force pushing ions across the membrane, due to the concentration gradient, can be counteracted by applying voltage *V*. The value $V_0$ at which the current becomes zero is known as the membrane or reversal potential and is given by the Nernst equation[21]

$$V_0 = (t_{Cl} - t_H)\,(k_B T/e)\,\ln(\Delta C) = -(2t_H - 1)\,(k_B T/e)\,\ln(\Delta C) \qquad (1)$$

where $k_B$ is the Boltzmann constant, $T$ is the temperature and $e$ is the elementary charge. If one of the transport numbers is unity, the other must be zero and then, it is said that a membrane displays perfect Nernst selectivity. Fig. 2a shows that for $\Delta C = 10$, the $I$-$V$ curves intersected the x-axis at the same $V$, which means that our membranes exhibited $V_0 \approx$ -58 mV, regardless of the absolute values of the HCl concentrations. This value is equal to $-(k_BT/e)\ln(\Delta C=10) \approx$ -58 meV at our measurement temperature of ~20 °C and, therefore, the observation implies $t_H \approx$ 1 or, equivalently, that all the ionic current through the membrane is due to proton transport. Within our experimental accuracy, the same perfect selectivity was also found for graphene (Supplementary Fig. 5).

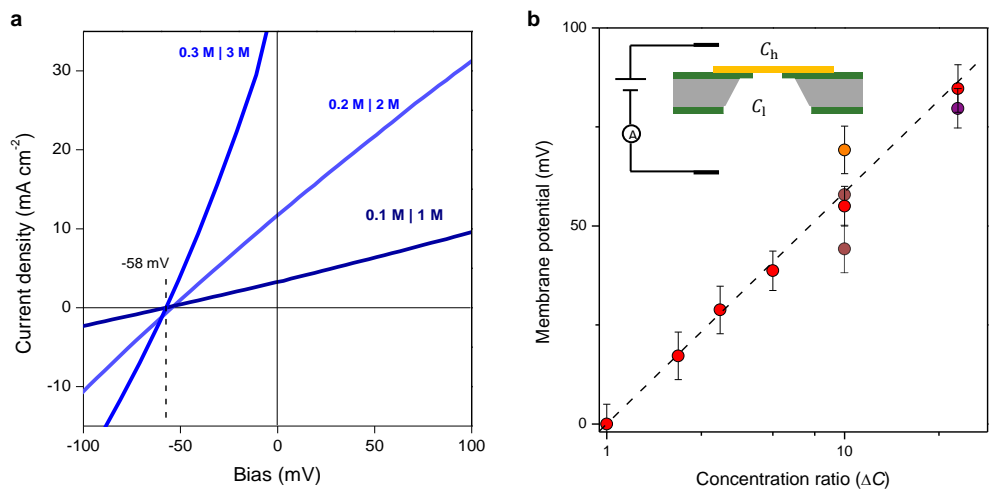

**Figure 2| Proton selectivity.** (**a**) Examples of $I$-$V$ characteristics for various HCl concentrations across a monolayer hBN membrane at a fixed $\Delta C = 10$. The current at zero voltage (intersection with the y-axis) was always positive. The reversal potential $V_0$ is given by the intersection of the $I$-$V$ curves with the x-axis and was $V_0 \approx$ -58 mV as marked by the dotted line. (**b**) $V_0$ for different $\Delta C$ and 4 different hBN devices (symbols of different color). Error bars, SD from different measurements. The black line is given by Eq. (1) for $t_H = 1$ and $t_{Cl} = 0$. Inset: Schematic of the experimental setup.

To corroborate the above result and obtain better statistics for the ion selectivity, we carried out similar measurements using different devices and several concentration ratios ranging from $\Delta C = 1$ to 30 (Fig. 2b). For all of them, we found membrane potentials consistent with the perfect proton selectivity in Eq. (1). The best fit to the data in Fig. 2b yields $t_H = 0.99 \pm 0.02$, or $t_H \approx$ 1. In control experiments, we verified our experimental approach using porous glass membranes. They allow large concentrations gradients but provide no ion selectivity because of large pore sizes. The latter experiments yielded $t_H = 0.81 \pm 0.04$ (Supplementary Fig. 6), in agreement with the transport numbers known for bulk hydrochloric acid ($t_H \approx$ 0.83, $t_{Cl} \approx$ 0.17)[20].

Finally, it is instructive to compare our results with those obtained previously in conceptually similar experiments but using CVD graphene[11]. The latter was reported to have $\sigma \approx$ 4 S cm$^{-2}$ at 1 M HCl, in clear disagreement with our experiments for mechanically-exfoliated graphene where $\sigma$ was nearly three orders of magnitude smaller. Furthermore, no current could be detected for 1 mM HCl concentration in our experiments; but large current densities of ~10 mA cm$^{-2}$ were reported in ref. [[11]] for CVD graphene membranes of the same area. The membrane potential reported for CVD graphene was also different, reaching only ~8 mV for $\Delta C =$ 10, or ~7 times smaller than what we found for our devices. All this shows that the ion transport properties of exfoliated 2D crystals are radically different from those of CVD films where atomic-scale defects and, possibly, even macroscopic ones[11] dominate ion transport. This conclusion is consistent with all the other evidence for intrinsic proton transport through 2D crystals, which was reported previously[1,2].

In conclusion, our experiments clearly demonstrate that mechanically-exfoliated, defect-free 2D crystals allow only proton transport and block even small ions such as chlorine that has one of the smallest hydrated diameters[19]. This provides further support to the view that the activation barriers found for proton transport through high-quality graphene and hBN do not involve vacancies and other atomic-scale defects[1] a conclusion important for further theory developments (e.g., for the hydrogenation model proposed in refs. [8,9]). Our results also have implications for the widely-discussed use of atomically-thin crystals as a novel platform for various separation technologies. In such technologies, selectivity is typically achieved by either perforating nanopores[22–25] or exploiting those naturally occurring in CVD films[26,27]. The fast permeation of H⁺ through the 2D bulk is usually ignored but can be important for designing and optimizing the membranes' properties.

# Perfect proton selectivity in ion transport through two-dimensional crystals

L. Mogg, S. Zhang, G.-P. Hao, K. Gopinadhan, D. Barry, B. L. Liu, H. M. Cheng, A. K. Geim, M. Lozada-Hidalgo

**Device fabrication and electrical measurements**

Device fabrication started by isolating atomically-thin layers of graphene and hBN from bulk crystals using the dry transfer technique[1]. The flake is first identified optically and then characterized using atomic force microscopy and Raman spectroscopy[1,2]. Supplementary Fig. 1 shows typical characterization data for one of the used hBN crystals. Similar characterization procedures were performed for graphene.

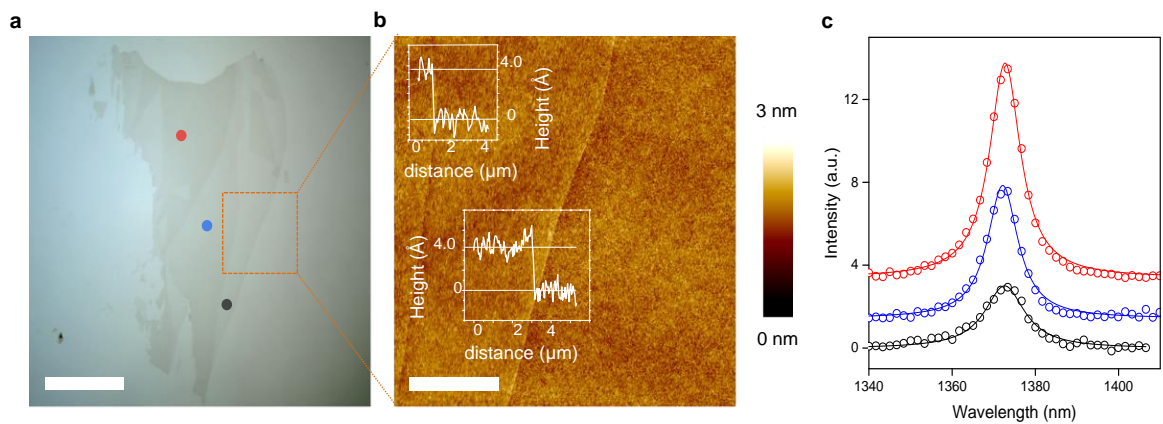

**Supplementary Figure 1 | Characterization of mono- and few- layer hBN. a,** Optical image of a typical hBN flake. Mono-, bi- and tri- layer regions are marked by black, blue and red dots, respectively. Scale bar, 15 μm. **b,** AFM image of the area marked in panel **a** by the red square. The insets show the step heights corresponding to the mono- and bi- layer regions. Scale bar, 4 μm. **c,** Raman spectra from the three areas marked by the dots in panel **a** (color coded). The solid lines are Lorentzian fits.

Supplementary Fig. 2 illustrates the device fabrication process. Several lithography, reactive ion etching and wet etching steps were performed to obtain a fully suspended SiN membrane with a 2-μm-diameter aperture in the center. The exfoliated 2D crystals were then suspended over the micro-fabricated apertures, following the recipe described in the previous report[3]. The crystals were then clamped down to the SiN substrate with a polymer washer. To this end, an SU-8 photo-curable epoxy washer was prefabricated with a 10-μm-diameter hole in the middle and transferred over the devices with its hole aligned with the aperture in the SiN membrane (Supplementary Fig. 2b). After the transfer, the seal was hard baked at 150 ˚C to ensure good adhesion to the substrate. For electrical measurements, the devices were clamped with O-rings and used to separate two reservoirs filled with HCl solutions; Ag/AgCl electrodes were placed inside each reservoir. Our measurement cell is illustrated in Supplementary Fig. 2g. *Keithley* 2636A SourceMeter was used to both apply voltage and measure current. The *I-V* characteristics were measured at voltages varying typically between ±200 mV and using sweep rates <0.1 V min⁻¹.

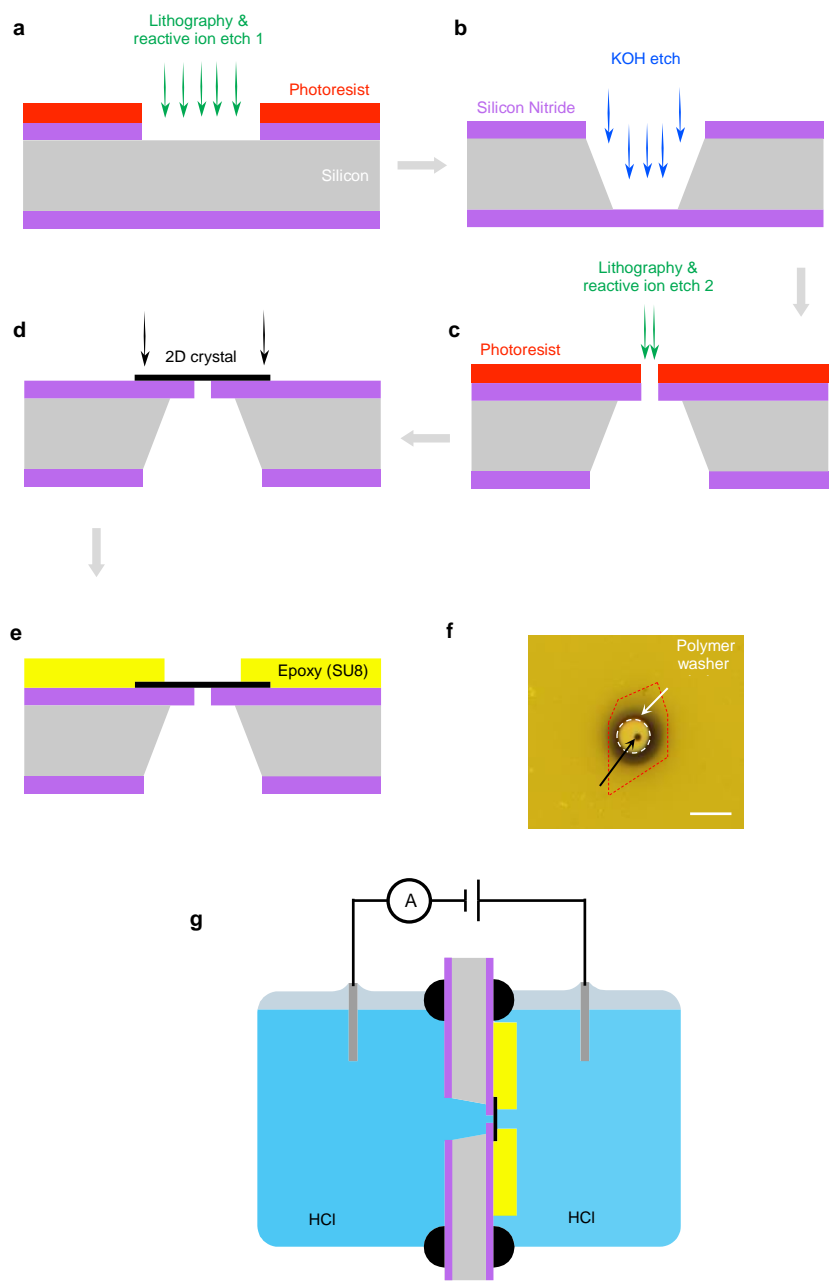

**Supplementary Figure 2| Experimental details**. **a-e**, Device fabrication flow. Arrows between panels indicate the order in which the different fabrication steps were performed. **f**, Optical micrograph of a final device (top view). The position of the 2D crystal is outlined by the red dotted curve; the circular aperture in SiN is marked with the black arrow; the hole in the polymer washer with the white dotted circle. Scale bar, 10 µm. **g**, Schematic of our liquid cell. The O-rings used to seal devices are represented with black circles.

To characterize our setup, we first determined typical leakage currents, in the absence of any proton conductive path. This was done in two different ways. First, a SiN substrate without an aperture was used to separate two HCl solution reservoirs. Second, a suspended 2D membrane device was used to separate two reservoirs filled with deionized water. In both cases only minute currents of the order of 1 pA were detected. This shows that electrical leakage provided little contribution to the obtained *I-V* characteristics of our 2D-membrane devices. Next, we characterized the maximum possible conductance through our apertures at a given HCl concentration. To this end, we measured devices in which the apertures in SiN were not covered with a 2D crystal (referred to as 'bare aperture devices'). Supplementary Fig. 3 shows that $\sigma$ of such devices scaled linearly with electrolyte concentration. Importantly, we found that for all concentrations, $\sigma$ of bare-aperture devices was $\gtrsim$1000 times larger than for those with a 2D-crystal membrane.

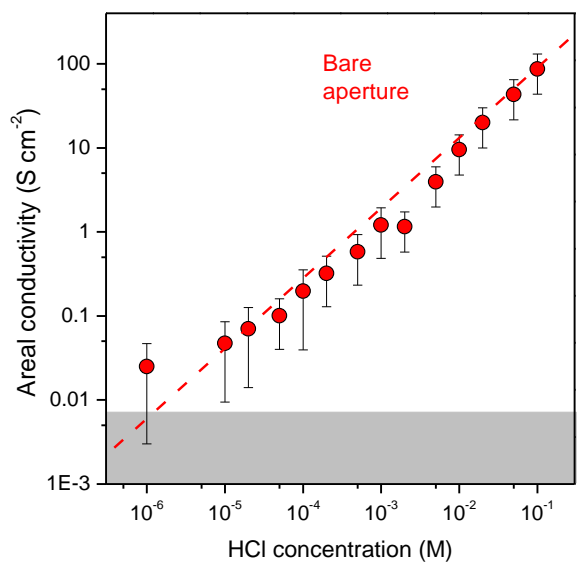

**Supplementary Figure 3| Conductance of bare-aperture devices.** Their $\sigma$ as a function of HCl concentration. Dotted line: Best linear fit to the data. The grey area indicates our detection limit determined by pA-range leakage currents.

**Leak tests using nanoballoons**

The most sensitive technique to detect microscopic defects in 2D crystals is gas-leak measurements using 'nanoballoons'[4,5]. In such experiments, a small (~1 μm³) microcavity in an oxidized Si wafer is sealed with a 2D crystal membrane and then filled with a chosen gas (typically, Ar) pressurized above 1 bar[4,5]. The pressure difference between the gas inside and outside the microcavity causes the 2D membrane to bulge upwards (inset Supplementary Fig. 4). It is possible to monitor changes in the gas pressure inside the microcavity by measuring the membrane deflection using AFM. In the absence of atomic-scale defects, the gas slowly leaks along the silicon oxide layer until the pressure inside and outside the chamber is equalized, a process that typically takes many hours. However, in the presence of even a single angstrom-sized defect (such as a vacancy), the pressure inside the microcavity equalizes typically in seconds[5,6].

To check that our membranes are defect-free, we carried out the above gas-leak experiments following the approach of refs. [4,5]. To this end, we etched microcavities in a Si/SiO₂ wafer and sealed them with monolayer graphene. The microcavities were pressurized by placing the devices inside a 'charging' chamber filled with Ar at 2 bar. After several days, the devices were then taken out of the charging chamber and their height profile was measured with AFM. Supplementary Fig. 4 shows typical results found for dozens of the membrane devices that were studied. The membranes were found to bulge upwards and the Ar leak rate was found to be ~10³ atoms per second, in agreement with permeability of the Si oxide layer[4]. Next, in control experiments, we intentionally introduced atomic scale vacancies by mild ultra-violet etch[5]. This procedure yields a defect density so low that it cannot be detected using Raman spectroscopy. Nevertheless, we found that the resulting nanoballoons did not inflate at all, even after leaving them in the charging chamber for over a month. This is consistent with rapid gas effusion through the 2D membranes such that angstrom-sized defects lead to their deflation within seconds, beyond time resolution of our approach[5,6]. The described experiments show that our mechanically-exfoliated crystals were defect-free, in agreement with the conclusions reached in refs. [5,6] for similar graphene devices.

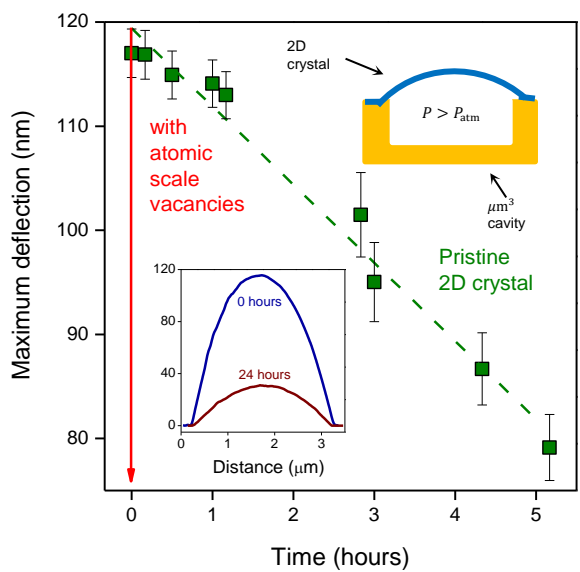

**Supplementary Figure 4I Leak tests using nanoballoons.** Maximum membrane deflection as a function of time. The data point at time zero hours corresponds to the first measurement after the device was taken out of the charging chamber. It normally took us only several minutes before the first data point was recorded. The red arrow indicates that in the presence of a few atomic-scale defects, we did not observe any bulging at all. Top inset: Schematic of our nanoballoons. Bottom inset: AFM traces taken through the center of an inflated nanoballoon at different times after taking it out of the charging chamber.

**Membrane potential measurements**

To measure the membrane potential for our membrane devices, they were placed to separate two reservoirs filled with HCl solutions at different concentrations. The membrane potential was measured by recording *I-V* characteristics and finding their intersection with the x-axis. Such intersection is known as the zero current or cell potential ($V_{cell}$) and has two components: the redox potential ($V_{redox}$) and the membrane potential ($V_o$):

$$V_{cell} = V_{redox} + V_o \qquad (1)$$

The redox potential appears due the electrodes' material and is independent of the studied membrane. Its value is well known for Ag/AgCl electrodes. For this reason, it is customary to remove this fixed contribution and report only $V_0$. We followed this convention. Nevertheless, to double-check this contribution, we also measured our devices using reference electrodes, instead of Ag/AgCl ones[7]. If using the reference electrodes, we indeed found $V_{cell} = V_0$, as expected.

**Selectivity of graphene devices**

The proton conductance through graphene membranes is at least $\sim$50 times lower than that for monolayer hBN. For this reason, parasitic capacitive contributions from the setup become significant and induce notable errors in the membrane potential measurements. To minimize this problem, we fabricated a device with many (nine) apertures (each of 2 μm in diameter) and then covered all nine with one large mechanically-exfoliated graphene monolayer. This was possible with graphene because, unlike hBN, it can be mechanically exfoliated into crystals of up to hundreds of microns across. Supplementary Fig. 5 shows *I-V* characteristics for this device when it was used to separate two HCl solutions at the concentration ratio $\Delta C = 10$. Hysteresis in the *I-V* curve was much smaller than for individual 2 μm apertures but still contributed towards the uncertainty in determining $V_0$, which was somewhat larger than that for our typical hBN devices (Supplementary Fig. 5). We obtained $V_0 = -55\pm9$ mV, which within the uncertainty corresponds to the perfect selectivity for protons.

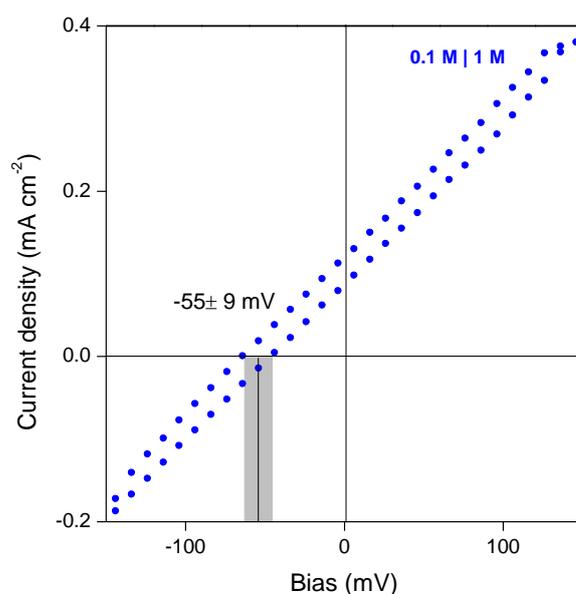

**Supplementary Figure 5| Proton selectivity for graphene.** *I-V* characteristics of a graphene-membrane device that separated two reservoirs with a concentration gradient of 10. The uncertainty in determining $V_0$ is marked by the grey rectangle.

**Bulk transport numbers for HCl**

As a reference, we carried out similar measurements of the membrane potential using a porous glass membrane. Supplementary Fig. 6 shows the values of $V_0$ extracted from these experiments.

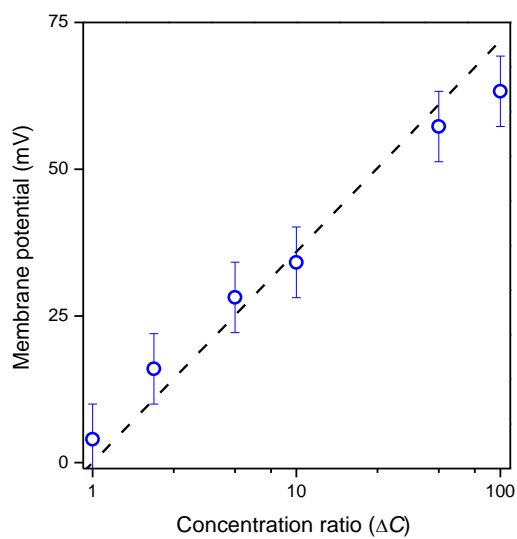

**Supplementary Figure 6| Reference measurements of the membrane potential for porous glass.**
Symbols: Our experimental data. The black line is given by Eq. (1) and the literature values $t_H$ = 0.83 and $t_{Cl}$ = 0.17 for bulk hydrochloric acid[8].